\documentclass[conference]{IEEEtran}
\IEEEoverridecommandlockouts
\usepackage{graphicx} 
\usepackage{amsmath}  
\usepackage{amssymb}
\usepackage{color}
\usepackage{multirow}
\usepackage{flushend}

\title{Reusing Softmax Hardware Unit for GELU Computation in Transformers}
\author{
\IEEEauthorblockN{Christodoulos Peltekis, Kosmas Alexandridis and Giorgos Dimitrakopoulos}
\IEEEauthorblockA{\small Electrical and Computer Engineering, Democritus University of Thrace, Xanthi, Greece\thanks{This work was supported by a research grant of Siemens EDA to Democritus University of Thrace for ``HLS Research for Systems-on-Chip''.}}}

\begin{document}

\maketitle

\begin{abstract}
Transformers have improved drastically the performance of natural language processing (NLP) and computer vision applications. The computation of transformers involves matrix multiplications and non-linear activation functions such as softmax and GELU (Gaussion Error Linear Unit) that are accelerated directly in hardware. Currently, function evaluation is done separately for each function and rarely allows for hardware reuse. To mitigate this problem, in this work, we map the computation of GELU to a softmax operator. In this way, the efficient hardware units designed already for softmax can be \emph{reused} for computing GELU as well. Computation of GELU can enjoy the inherent vectorized nature of softmax and produce in parallel multiple GELU outcomes. Experimental results show that computing GELU via a pre-existing and incrementally modified softmax hardware unit (a) does not reduce the accuracy of representative NLP applications and (b) allows the reduction of the overall hardware area and power by 6.1\% and 11.9\%, respectively, on average.
\end{abstract}

\section{Introduction}
Transformers are a deep learning model used for natural language processing~\cite{llama, t5} and computer vision tasks~\cite{vit}. They utilize ``self-attention'' mechanism to process sequential input data. Transformers can process the entire input data capturing context and relevance.

\begin{figure}
\centering
\includegraphics[width=0.6\columnwidth]{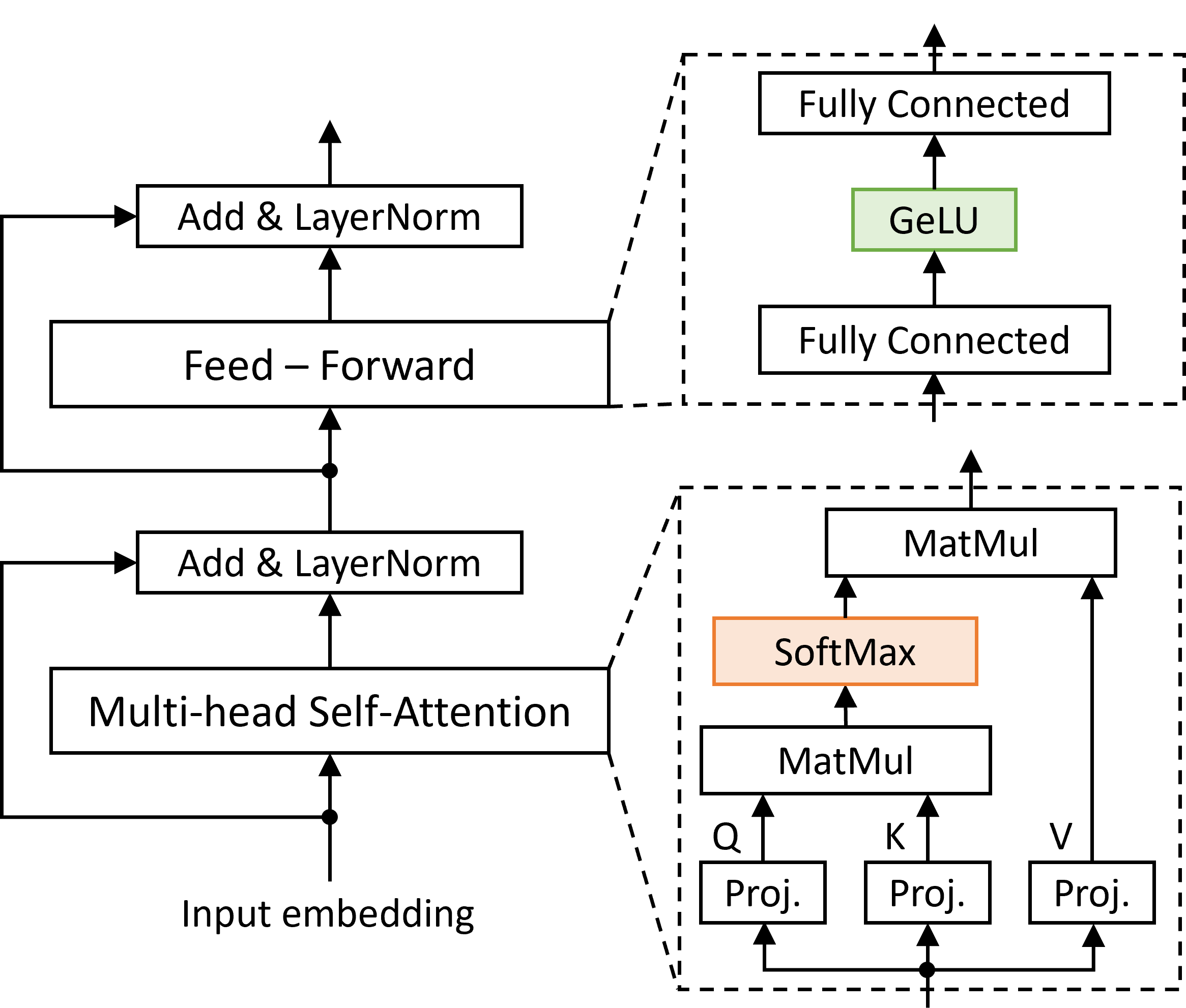}
\caption{An encoder-only transformer layer. For instance, BERT-base~\cite{bert} consists of twelve such layers.}
\label{f:tranformer}
\end{figure}

Transformer networks are composed of encoder and decoder blocks~\cite{attention-is-all-you-need} that include mainly matrix multiplications as well as softmax, normalization and GELU operations. Fig.~\ref{f:tranformer} depicts a layer of an encoder-only transformer~\cite{bert,albert}.  
The input embedding is first projected to Query ($Q$), Key ($K$) and Value ($V$) matrices through a linear transformation. Then, $Q$ and $K$ matrices are multiplied and scaled to calculate, for each embedding, the importance of its neighbors. The result is passed through a softmax operator and the output is multiplied with matrix $V$ to compute the attention matrix. To complete self-attention the output is normalized and added to the input of the attention block.
The self-attention block is followed by a feed-forward block that consists of two fully-connected layers that are separated by a GELU activation function. 
Decoder and encoder blocks follow a similar structure. Their difference is that the decoder consists of two self-attention blocks followed by a feed-forward block~\cite{attention-is-all-you-need}.

In transformers, the time spent for function evaluation, such as softmax, is not negligible~\cite{softermax}. Therefore, to mitigate this performance overhead, transformer accelerators are equipped with specialized softmax hardware units~\cite{bo-yuan}\nocite{asap2020, edgebert, softermax, kouretas-softmax,koca, somalib}--\cite{softmax-dac}. Although the available state-of-the-art softmax hardware units adopt different architectures, they all share an inherently parallel operation:
They receive an input vector of data elements and compute in parallel an output vector of probabilities according to the definition of softmax.

In this work, our goal is to \emph{leverage} state-of-the-art softmax architectures and their inherent parallelism for computing other activation functions such as GELU (Gaussion Error Linear Unit). To achieve this goal, we first map the computation of GELU to a softmax operator. Then, we show how to incrementally modify a state-of-the-art softmax hardware unit to a dual-mode unit that computes both GELU and softmax functions. 
Most importantly, we guarantee that the computation of GELU can enjoy the inherent vectorized (parallel) nature of softmax and compute multiple GELU outcomes in parallel.

To quantify the accuracy of computing GELU via softmax, we computed inference on various BERT-based applications~\cite{mnli}\nocite{CoLA, sst2, mrpc}--\cite{rte} using three GELU variants. The proposed one, a fully accurate one that uses floating point arithmetic and a state-of-the-art integer GELU~\cite{ibert}. In all cases, the proposed approach achieves the same accuracy as compared to the rest models.

To assess hardware complexity we implemented the proposed combined design and compared it to the integer GELU~\cite{ibert} assuming the same degree of available parallelism. The combined GELU-softmax hardware unit leads to more area and power efficient designs than the ones that compute softmax and GELU separately. The savings achieved are 3.8\%--8.4\% for area and 10.7\%--13.2\% for power.

\section{GELU computation via a softmax operator}
Softmax transforms a vector of $N$ values $\vec{x} = [x_1, x_2, \ldots, x_n]$ to a vector of probabilities. The $i$th element of the resulting output vector $\vec y = \text{softmax}^N (\vec{x})$ equals
\begin{equation}
y_i = \text{softmax}^N_{i}(\vec{x}) = \frac{e^{x_i}}{\sum^N_{j=1} e^{x_j}}
\label{e:softmax}
\end{equation}

\noindent In its simplest form, the first element of the output of softmax when applied on a two-element vector $[x_1, x_2]$ (i.e., $N=2$), equals 
\begin{equation}
\text{softmax}^2_1([x_1, x_2]) = \frac{e^{x_1} } { e^{x_1} + e^{x_2} }
\label{e:softmax-two}
\end{equation}

GELU follows a different form and is computed via the error function as follows~\cite{gelu}:
\begin{equation}
\text{GELU}(z) = \frac{1}{2} z \left (1 + \text{erf}\left(\frac{z}{\sqrt{2}}\right) \right)    
\label{e:orig-gelu}
\end{equation}

In~\cite{gelu} it was shown that GELU can be approximated via the $\tanh$ function as follows:
\begin{equation}
\text{GELU}(z)\!=\!  \frac{1}{2} z \left ( 1 + \tanh \left [ \sqrt{\frac{2}{\pi}} \left (z + 0.044715 z^3\right ) \right ]\right ) 
\label{e:approx-gelu}
\end{equation}

\noindent Setting $k= \sqrt{\frac{2}{\pi}} \left (z + 0.044715 z^3 \right)$, we get that
\begin{equation}
\text{GELU}(z) = \frac{1}{2}\,z\, \bigl( 1 + \tanh(k)  \bigr )
\label{e:gelu-2}
\end{equation}

Expressing $\tanh$ function in exponential form, we can write the term in the parenthesis of~\eqref{e:gelu-2} as follows:
\begin{equation}
1 + \tanh(k) = 1 + \frac{e^k - e^{-k}}{e^k + e^{-k}} = \frac{2e^k}{e^k + e^{-k}}
\label{e:tanh-1}
\end{equation}

\noindent By replacing~\eqref{e:tanh-1} in~\eqref{e:gelu-2}, we effectively replace $\tanh$ from the approximate computation of GELU~\eqref{e:approx-gelu} with a fraction of exponentials:
\begin{equation}
\text{GELU}(z) = z\,\left( \frac{e^k}{e^k + e^{-k}} \right)    
\label{e:temp}
\end{equation}
The fraction of exponentials in~\eqref{e:temp} has exactly the same form as the first output element of softmax when applied on a two-element vector $[k, -k]$ (see Eq.~\eqref{e:softmax-two}). Following this observation, 
we can compute GELU via a two-element softmax operator
as follows:
\begin{equation}
\text{GELU}(z) = z\, \text{softmax}^2_1\bigl([k, -k] \bigr )
\label{e:gelu-with-softmax}
\end{equation}

To minimize cost, we aim at computing the two-element softmax (i.e., softmax$^2$) needed in~\eqref{e:gelu-with-softmax} by 
\emph{reusing a generic softmax hardware unit} that operates on $N$ input values $[x_1, x_2, \ldots, x_n]$. 
This goal can be satisfied in two ways:
\begin{itemize}
\item  
Set $x_1 = k$, $x_2=-k$, and the rest $x_i$, with $i > 2$ equal to the minimum possible value. Keep only the first output element $y_1$ of softmax.
\item
Modify the $N$-element softmax hardware operator to operate independently on $N/2$ two-element subvectors. 
\end{itemize}

In this work, we follow the second approach and show how a representative softmax hardware operator can be transformed to $N/2$ independent softmax operators that each one operates on two-element subvectors. 
As shown in the next Section~\ref{s:hardware} this choice also simplifies the parallelization of GELU computations.

\section{Combined GELU-Softmax Hardware Unit}
\label{s:hardware}
The reuse of a softmax hardware unit for computing GELU according to~\eqref{e:gelu-with-softmax} requires two hardware modifications. The first one refers to the extra hardware needed ``around'' softmax for computing $k$ for each input and for finalizing the GELU result after multiplying the input with the result of the two-element softmax. The second one involves the modifications needed in softmax itself to operate in parallel and independently across multiple two-element vectors of the form $[k, -k]$. 

\subsection{Softmax with configurable vector width}
Softmax~\eqref{e:softmax} receives a vector of $N$ input values and returns a vector of $N$ probabilities. To have a stable result, the maximum of all elements of $\vec x$ is identified first and subtracted from the rest values of $\vec x$:
\begin{equation}
y_i = \text{softmax}^N_i(\vec{x}) = \frac{e^{x_i-\max\vec x}}{\sum^N_{j=1} e^{x_j-\max\vec x}}
\label{e:softmax-max}
\end{equation}

\noindent In this way, the powers seen at each exponent $e^{x_i - \max \vec x}$ refer to zero or negative values. This simplifies the approximation of the exponents that can be performed either through table lookup, polynomial approximations or piece-wise linear approaches.

The next step after approximating input exponents $e^{x_i - \max \vec x}$, is the addition of the exponents of all input elements ${\sum^N_{i=1} e^{x_i-\max\vec x}}$. This can be done either using a tree of adders~\cite{bo-yuan, asap2020, edgebert} or following an online approach~\cite{online-softmax, softermax} that computes this sum in parallel to the identification of the maximum value of $\vec x$ needed in the first step.

The last step for computing softmax involves the normalization of the input exponents, where each input exponent is divided by the sum of all other exponents. This last step is either performed using a divider unit per vector element~\cite{softermax}\footnote{The number of dividers can be less than the output vector size depending on the output throughput requirements~\cite{benini-islped23}.}, or performing computation in the logarithm domain where division is turned into a subtraction~\cite{bo-yuan}. 

In this work, following the architecture of~\cite{bo-yuan, asap2020, edgebert}, we perform division in the logarithm domain. Thus, we consider for softmax the following implementation:
\begin{equation}
y_i = \text{softmax}^N_i(\vec{x}) = 
e^{x_i - \max \vec x - \log \bigl(\sum_j e^{x_j - \max \vec x} \bigr )}
\label{e:softmax-log-domain}
\end{equation}
For each exponentiation step we adopt the piecewise linear approximations (PWL) similar to~\cite{zhu2020efficient}. To do this, each $e^{x-\max \vec x}$ is written as $2^{(x-\max \vec x)\log_2 e} = 2^u\cdot 2^v$, where $u$ is the integer part of the number and $v$ is the fraction~\cite{koca, zhu2020efficient}.
Integer powers of $2$ such as $2^u$ are implemented as shifts, while $2^{v}$ is computed as an eight-piece PWL approximation on the range $[0, 1)$.
The coefficients are derived using~\cite{python-pwl} on the said range. For computing the logarithm of the sum of exponents, we use exactly the same architecture as the PWL forward logarithm converter of~\cite{logarithm-kaist}.

\begin{figure}
\centering
\includegraphics[width=0.9\columnwidth]{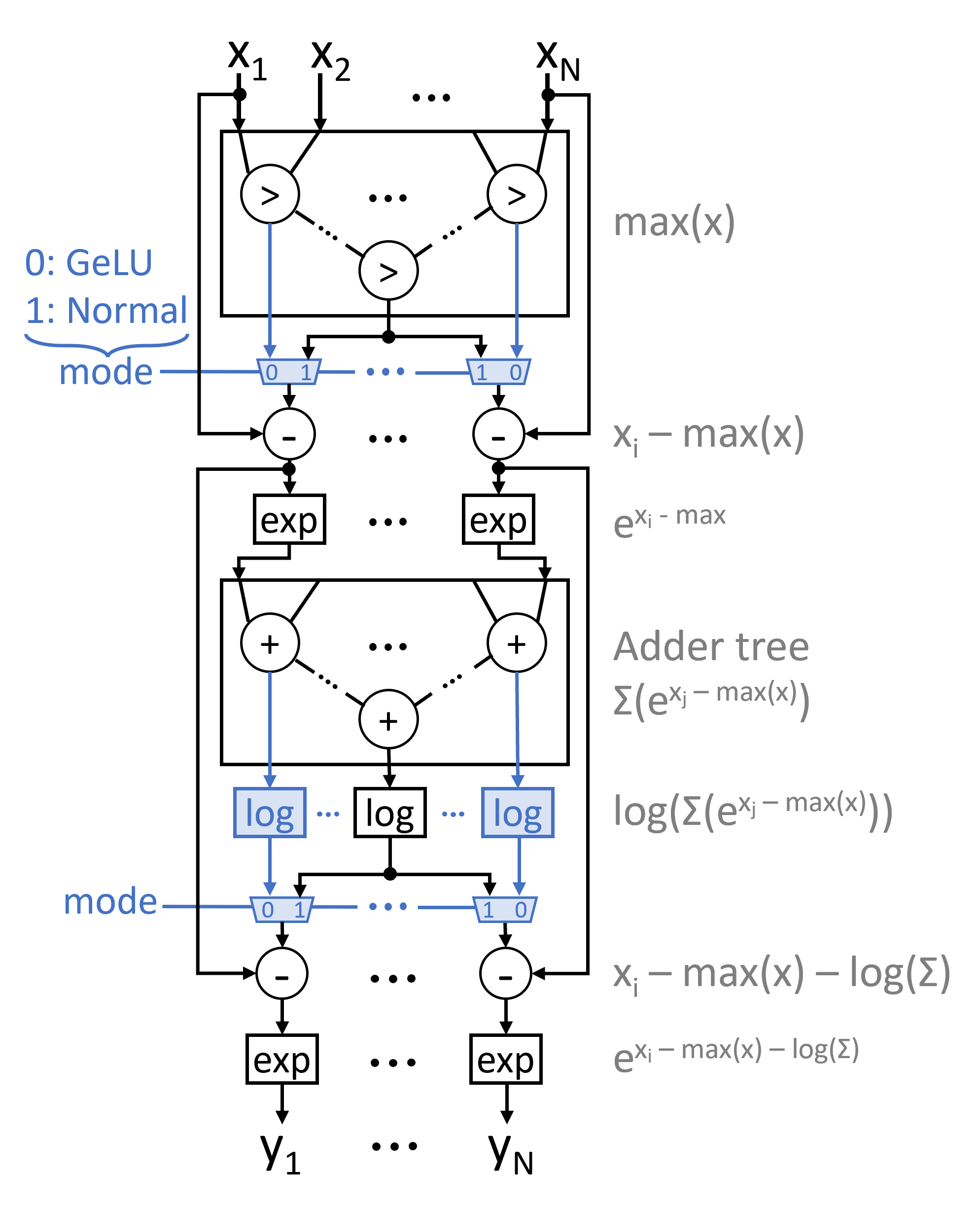}
\caption{The dual-mode softmax hardware unit. Computation follows Eq.~\eqref{e:softmax-log-domain} that implements division in the logarithm domain. The extra logic required for supporting the dual-mode of operation is highlighted in blue.}
\label{f:dual-mode-softmax}
\end{figure}

In this work, we need a configurable softmax module that operates in two different modes. When in normal mode softmax operates in all $N$ input elements at once. On the contrary, when in GELU mode, it operates independently on $N/2$ two-element vectors thus maximizing parallelism. To achieve this property we need to apply three incremental changes to a softmax module that computes~\eqref{e:softmax-log-domain}:

1) When in normal mode, the module that identifies the maximum of all input elements should return one output. In GELU mode, the same module should return one maximum element for every pair of elements ($N/2$ in total). This feature adds only a minimum multiplexing cost, since computing the global maximum of a vector already involves comparing locally all pairs of vector elements. Then, according to the chosen mode, the appropriate maximum value is subtracted from the input elements.

2) The adder tree that computes the sum of all exponents should return both the final sum (when in normal mode), as well as, the $N/2$ sums of consecutive pairs (when in GELU mode). This functionality is readily available in the design, since any adder tree computes the sum of consecutive pairs at the first level of the tree and the full-length sum at the last level of addition.

3) The computed sums are driven to separate logarithm unit. In normal mode, only one logarithm unit is needed for the whole softmax. On the contrary, when in GELU mode, $N/2$ logarithms are computed independently. 
Besides necessary multiplexing logic, these logarithm units are the only extra hardware needed for designing this dual-mode softmax unit.

To complete computation of~\eqref{e:softmax-log-domain} the output of each logarithm unit is subtracted from the corresponding difference $x_i - \max \vec x$ that is already available in both operating modes. At the end, to return from the logarithm domain, the result of each subtraction is passed to a separate exponent unit. 

Fig.~\ref{f:dual-mode-softmax} highlights the overall architecture of the dual-mode softmax unit that implements the required changes (steps 1--3). The experimental results presented in the next Section IV show that the transformation of a state-of-the-art single-mode softmax unit~\cite{edgebert} to a dual-mode one, incurs only marginal costs in area and power. 

\begin{figure}
\centering
\includegraphics[width=0.9\columnwidth]{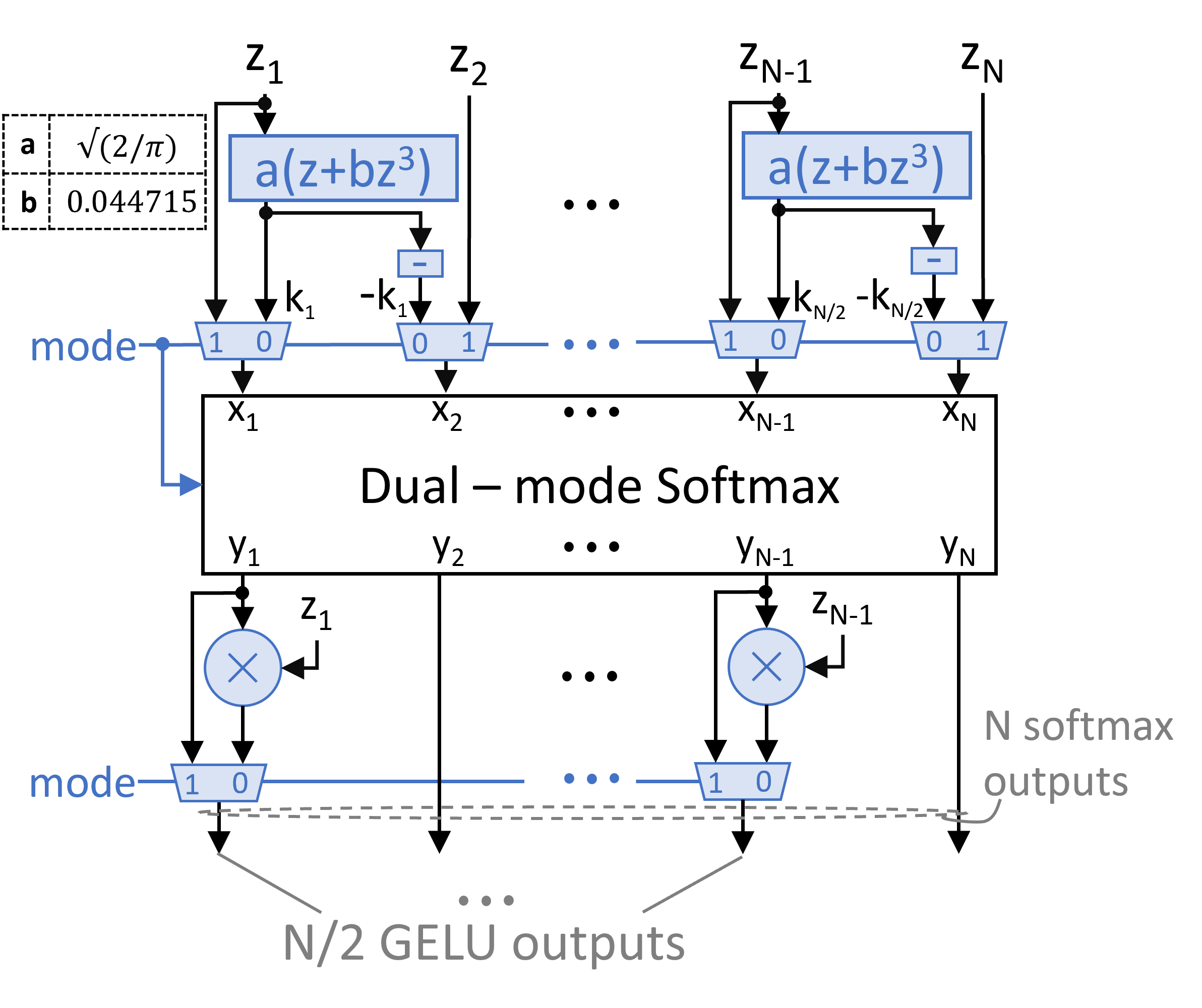}
\caption{The combined GELU/softmax hardware unit. When in normal mode, softmax is driven by $N$ inputs $z_i$ and produces $N$ output probabilities. When in GELU mode, half of the inputs and outputs are used to compute $N/2$ GELU outcomes in parallel.}
\label{f:gelu-implementation}
\end{figure}

\subsection{Computing multiple GELUs using dual-mode softmax unit}
The dual-mode softmax unit allows us to compute independently the softmax of $N/2$ two-element vectors. We leverage this property \emph{to compute $N/2$ GELU outcomes in parallel}.
The organization of this combined GELU-softmax unit is depicted in Fig.~\ref{f:gelu-implementation}.

The $N/2$ inputs $z_i$, with an odd index $i$, are passed to a datapath of multipliers and adders that computes the corresponding values $k_i$ and $-k_i$, respectively. Then, the $N/2$ two-element vectors $[k_i, -k_i]$ are given to the softmax unit that is configured to operate in GELU mode.
Following~\eqref{e:gelu-with-softmax}, to finish the computation of each $\text{GELU}(z_i)$, we need to multiply each $z_i$ with the first output produced by softmax for each two-element subvector.
In this way, we are able to compute multiple GELU outcomes in parallel using extra multipliers and adders around \emph{a shared dual-mode softmax unit}.

\begin{table*}[t!]
    \centering
    \caption{Performance of the approximate GELU functions on BERT model for 8 tasks of the GLUE dataset~\cite{glue}. }
    \begin{tabular}{|c|c||c|c|c|c|c|c|c|c|}
        \hline
          \multicolumn{2}{|c||}{}     &{\bf STS2}&{\bf MNLI-m}& {\bf MNLI-mm}&{\bf QQP}&{\bf QNLI}&{\bf CoLA}&{\bf MRPC}& {\bf RTE}  \\\hline \hline
        \multirow{3}{*}{Accuracy (\%)}
        &FP32    & 92.7 & 79.1   & 84.2    &90.8 & 94.1 & 72.4 & 85.8 & 73.8 \\
        &i-GELU  & 92.4 & 79.0   & 84.0    &90.7 & 94.1 & 72.1 & 86.6 & 73.7 \\
        &Proposed& 92.5 & 79.1   & 84.1    &90.7 & 94.0 & 72.3 & 85.7 & 73.8 \\\hline \hline
        Mean Absolute &i-GELU  &0.054 & 0.0824& 0.094   &0.098  & 0.064 & 0.076 & 0.1204 & 0.179 \\
        Error         &Proposed&0.0039& 0.0057& 0.0061  &0.0083 &0.0045 &0.0067 & 0.0109 & 0.0154 \\\hline
    \end{tabular}
    \label{t:accuracy}
\end{table*}

\section{Experimental results}
The experimental evaluation is twofold: First, we explore the accuracy of computing GELU using a dual-mode softmax unit in representative BERT-based applications~\cite{mnli}\nocite{CoLA, sst2, mrpc}--\cite{rte}. Second, we quantify the savings achieved by 
using the combined GELU-softmax hardware unit shown in Fig.~\ref{f:gelu-implementation} relative to state-of-the-art implementations that employ independent designs for GELU and a single-mode softmax unit.
For all designs under comparison, we assume 16-bit fixed-point inputs with five integer bits and
32-bit integer arithmetic for all internal operations similar to the one employed for i-GELU in~\cite{ibert}. 

\subsection{Accuracy of GELU function in BERT applications}
\label{s:apps}
We evaluate the \textbf{accuracy of the proposed GELU computation} by executing inference on the BERT model~\cite{bert} using  eight tasks of the GLUE dataset~\cite{glue}.
For comparisons, we considered three transformer implementations: (a) Model 'FP32' computes {\bf all operations} including matrix multiplications and function evaluation using {\bf single-precision} floating point arithmetic; (b) Model 'i-GELU' employs single-precision floating point arithmetic for all parts of the transformer \textbf{except for the GELU function}, where computations are done in integer arithmetic according to~\cite{ibert}; (c) The 'Proposed' model that employs floating point arithmetic for all parts of the transformer \textbf{except for GELU and softmax} functions that are computed using the proposed combined integer GELU-softmax unit.

The results gathered are depicted in Table~\ref{t:accuracy}. In terms of inference accuracy the performance of all three methods is indistinguishable.
To have a more clear insight on the arithmetic performance of the proposed design and `i-GELU' relative to `FP32', we measured the mean absolute error of the outputs of the model in those two cases and compared it to that of `FP32'. In all cases, the proposed design that follows the original $\tanh$ approximation that is computed via softmax, exhibits a smaller error that ranges between $10^{-2}$ -- $10^{-3}$ for the various applications.

\subsection{Hardware Complexity Evaluation}
The hardware implementation of the proposed module has two goals. First, we aim at quantifying the cost of transforming a state-of-the-art single-mode softmax hardware unit~\cite{edgebert} to a dual-mode one that could operate on two different vector widths. Second, we need to assess how much hardware area and power is saved when using the combined GELU-softmax unit relative to state-of-the-art designs that use separate hardware modules for GELU and softmax.

All designs that are open-sourced in Github~\cite{git} were implemented in C++ and synthesized to Verilog RTL using Catapult HLS for a target clock frequency of 500 MHz driven by a 45-nm standard-cell library. The final timing/area results were derived from the Oasys logic synthesis tool. The power consumption was estimated after synthesis using the PowerPro power analysis and optimization tool. The input data used for power estimation come from traces derived by the applications examined in Section~\ref{s:apps}.

\begin{table}[h!]
\centering
\caption{The area and power of a single-mode and a dual-mode softmax unit for two vector widths.}
\begin{tabular}{|c||c|c|c||c|c|c|} 
\hline
\multirow{3}{*}{N} &
\multicolumn{3}{|c||}{Area ($ \mu m^2$)} & \multicolumn{3}{|c|}{Power ($mW$)} \\ \cline{2-7}
& single & dual & \multirow{2}{*}{Diff.} & single & dual & \multirow{2}{*}{Diff.} \\ 
& mode   & mode &       & mode   & mode & \\
\hline
8  & 92309 & 100675	 & 9.1\% & 13.51 & 13.95 & 3.3\%\\
32 & 363703	& 403025 & 10.8\% & 57.74	& 58.84	& 1.9\%  \\ \hline
\end{tabular}
\label{t:ap_augmented}
\end{table}

Table~\ref{t:ap_augmented} depicts the area and power results of a single-mode and a dual-mode softmax unit for two input sizes, e.g., $N=8$ and $N=32$. The results show that the \textbf{hardware cost for enhancing the functionality of a  state-of-the-art single-mode softmax} unit~\cite{bo-yuan, asap2020, edgebert} is marginal and scales well for both small and large input vector sizes. The average area and power overhead of adding an extra mode of operation to softmax for the two examined vector sizes is 9.9\% and 2.6\%, respectively.

\begin{figure}[ht!]
\centering
\includegraphics[width=0.95\columnwidth]{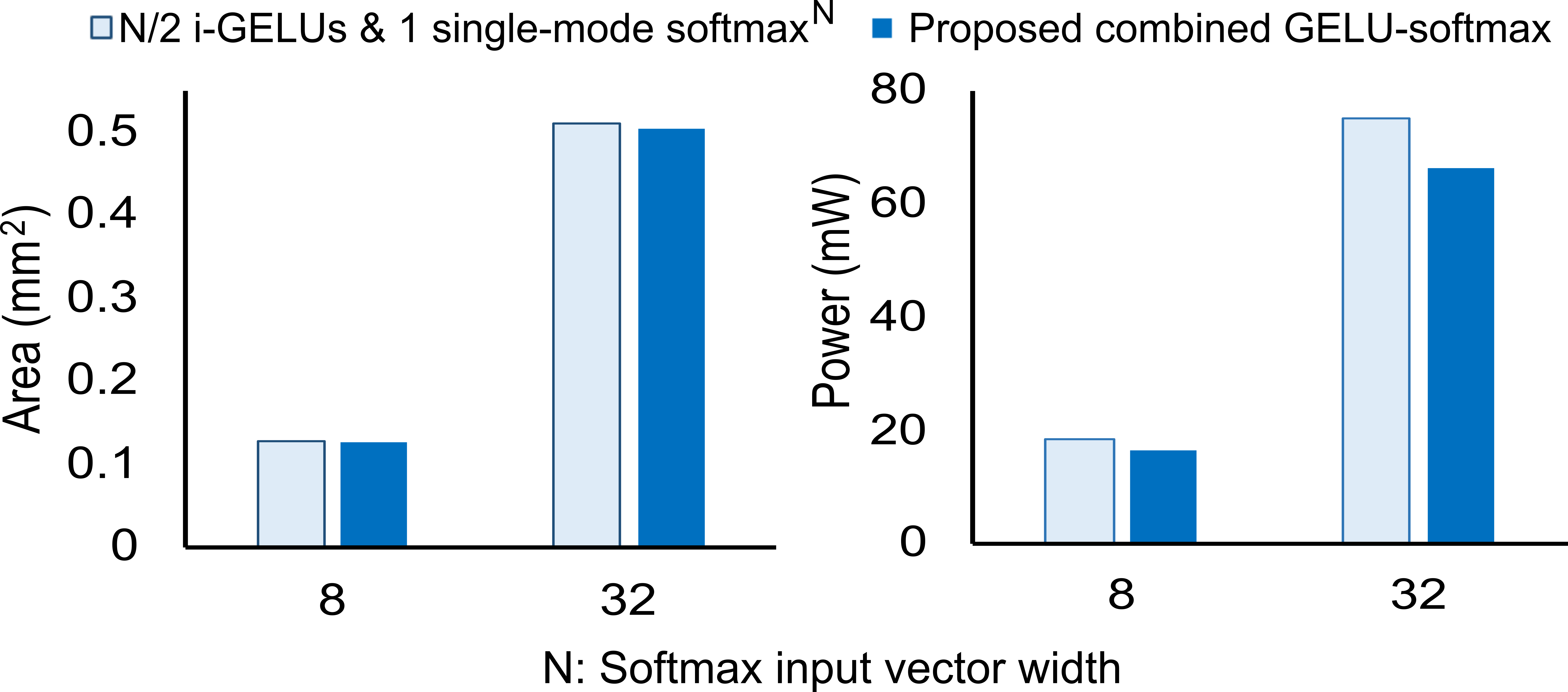}
\caption{The area and power of the combined GELU-softmax unit introduced in this work and the design that employs $N/2$ i-GELU~\cite{ibert} units and a state-of-the-art single-mode softmax unit~\cite{edgebert}. The dual-mode softmax unit used in the proposed design is an enhanced derivative of the same single-mode softmax unit.}
\label{f:comp_area}
\end{figure}

To quantify the textbf{area-power savings earned when following the combined GELU-softmax} architecture, we performed the following comparison. We designed \textbf{a combined GELU-softmax unit} for $N=8$ and $N=32$ following the organization shown in Fig.~\ref{f:gelu-implementation} in each case. This unit can either produce $N$ output elements of a softmax$^N$ operation, or it can produce in parallel $N/2$ GELU outcomes. To have a fair comparison, we compared this architecture with a design that uses \textbf{$N/2$ i-GELU units}~\cite{ibert} that compute multiple GELUs in parallel and \textbf{one single-mode softmax unit}~\cite{bo-yuan, asap2020, edgebert}. The single-mode softmax unit, used in comparisons, is the same design from which the dual-mode softmax unit, used in the proposed design, was derived. By using $N/2$ i-GELU units, this alternative design matches the computation throughput of the proposed design.

Fig.~\ref{f:comp_area} depicts the area and power of both designs under comparison. 
For $N=8$, the proposed unit requires 8.4\% less area and 10.7\% less power. While for $N=32$ elements, we report a 3.8\% decrease in area and 13.2\% decrease in power. In both cases, the majority of the power and area is due to the softmax units used in each case. The power reductions observed are a result of the smaller dynamic power used by the datapath used 'around' the dual-mode softmax relative to the polynomial evaluations employed in i-GELU~\cite{ibert}.

\section{Conclusions}
GELU and softmax are two critical components of modern transformer architectures and their efficient acceleration in hardware is of paramount importance. In this work, we leverage the benefits of state-of-the-art softmax architectures and their inherent vector-parallel operation to design efficient vector-parallel combined GELU-softmax units. This combined operation made possible by (a) a new mathematical transformation that computes GELU using a two-element softmax operator and additional multiplication and additions, as well as (b) the low-cost design of a softmax operator with configurable vector width.
The proposed combined GELU-softmax hardware unit not only reduces area and power consumption but also offers the same accuracy for representative NLP applications.

\bibliographystyle{IEEEtran}
\bibliography{refs}

\begin{thebibliography}{10}
\providecommand{\url}[1]{#1}
\csname url@samestyle\endcsname
\providecommand{\newblock}{\relax}
\providecommand{\bibinfo}[2]{#2}
\providecommand{\BIBentrySTDinterwordspacing}{\spaceskip=0pt\relax}
\providecommand{\BIBentryALTinterwordstretchfactor}{4}
\providecommand{\BIBentryALTinterwordspacing}{\spaceskip=\fontdimen2\font plus
\BIBentryALTinterwordstretchfactor\fontdimen3\font minus \fontdimen4\font\relax}
\providecommand{\BIBforeignlanguage}[2]{{%
\expandafter\ifx\csname l@#1\endcsname\relax
\typeout{** WARNING: IEEEtran.bst: No hyphenation pattern has been}%
\typeout{** loaded for the language `#1'. Using the pattern for}%
\typeout{** the default language instead.}%
\else
\language=\csname l@#1\endcsname
\fi
#2}}
\providecommand{\BIBdecl}{\relax}
\BIBdecl

\bibitem{llama}
H.~Touvron, T.~Lavril, G.~Izacard, X.~Martinet, M.-A. Lachaux, T.~Lacroix, B.~Rozi{\`e}re, N.~Goyal, E.~Hambro, F.~Azhar \emph{et~al.}, ``Llama: Open and efficient foundation language models,'' \emph{arXiv preprint arXiv:2302.13971}, 2023.

\bibitem{t5}
C.~Raffel, N.~Shazeer, A.~Roberts, K.~Lee, S.~Narang, M.~Matena, Y.~Zhou, W.~Li, and P.~J. Liu, ``Exploring the limits of transfer learning with a unified text-to-text transformer,'' \emph{The Journal of Machine Learning Research}, vol.~21, no.~1, pp. 5485--5551, 2020.

\bibitem{vit}
A.~Dosovitskiy, L.~Beyer, A.~Kolesnikov, D.~Weissenborn, X.~Zhai, T.~Unterthiner, M.~Dehghani, M.~Minderer, G.~Heigold, S.~Gelly, J.~Uszkoreit, and N.~Houlsby, ``An image is worth 16x16 words: Transformers for image recognition at scale,'' in \emph{Intern. Conf. on Learning Representations (ICLR)}, 2021.

\bibitem{attention-is-all-you-need}
A.~Vaswani, N.~Shazeer, N.~Parmar, J.~Uszkoreit, L.~Jones, A.~N. Gomez, {\L}.~Kaiser, and I.~Polosukhin, ``Attention is all you need,'' \emph{Advances in neural information processing systems}, vol.~30, 2017.

\bibitem{bert}
J.~Devlin, M.-W. Chang, K.~Lee, and K.~Toutanova, ``{BERT}: Pre-training of deep bidirectional transformers for language understanding,'' in \emph{North American Chapter of the Association for Computational Linguistics}, 2019.

\bibitem{albert}
Z.~Lan, M.~Chen, S.~Goodman, K.~Gimpel, P.~Sharma, and R.~Soricut, ``{ALBERT}: A lite bert for self-supervised learning of language representations,'' in \emph{Intern. Conf. on Learning Representations (ICLR)}, 2020.

\bibitem{softermax}
J.~R. Stevens, R.~Venkatesan, S.~Dai, B.~Khailany, and A.~Raghunathan, ``Softermax: Hardware/software co-design of an efficient softmax for transformers,'' in \emph{Design Automation Conference (DAC)}, 2021, pp. 469--474.

\bibitem{bo-yuan}
B.~Yuan, ``Efficient hardware architecture of softmax layer in deep neural network,'' in \emph{{IEEE} Intern. System-on-Chip Conference (SOCC)}, 2016.

\bibitem{asap2020}
Z.~Wei, A.~Arora, P.~Patel, and L.~John, ``Design space exploration for softmax implementations,'' in \emph{{IEEE}, Intern. Conf. on Application-specific Systems, Architectures and Processors (ASAP)}, 2020.

\bibitem{edgebert}
T.~Tambe, C.~Hooper, L.~Pentecost, T.~Jia, E.-Y. Yang, M.~Donato, V.~Sanh, P.~Whatmough, A.~M. Rush, D.~Brooks, and G.-Y. Wei, ``{EdgeBERT}: Sentence-level energy optimizations for latency-aware multi-task nlp inference,'' in \emph{{IEEE/ACM} Intern. Symp. on Microarchitecture (MICRO)}, 2021, p. 830–844.

\bibitem{kouretas-softmax}
I.~Kouretas and V.~Paliouras, ``Hardware implementation of a softmax-like function for deep learning,'' \emph{Technologies}, vol.~8, p.~46, 2020.

\bibitem{koca}
N.~A. Koca, A.~T. Do, and C.-H. Chang, ``Hardware-efficient softmax approximation for self-attention networks,'' in \emph{{IEEE} Intern. Symposium on Circuits and Systems (ISCAS)}, 2023.

\bibitem{somalib}
H.~C. Prashanth and M.~Rao, ``Somalib: Library of exact and approximate activation functions for hardware-efficient neural network accelerators,'' in \emph{IEEE Intern. Conf. on Computer Design (ICCD)}, 2022, pp. 746--753.

\bibitem{softmax-dac}
J.~Kim, J.~Lee, J.~Choi, J.~Han, and S.~Lee, ``Range-invariant approximation of non-linear operations for efficient bert fine-tuning,'' in \emph{Design Automation Conference (DAC)}, 2023.

\bibitem{mnli}
A.~Williams, N.~Nangia, and S.~R. Bowman, ``A broad-coverage challenge corpus for sentence understanding through inference,'' in \emph{Intern. Conf. of the North American Chapter of the Association for Computational Linguistics}, 2018.

\bibitem{CoLA}
A.~Warstadt, A.~Singh, and S.~R. Bowman, ``Neural network acceptability judgments,'' \emph{Transactions of the Association for Computational Linguistics}, vol.~7, pp. 625--641, 2019.

\bibitem{sst2}
R.~Socher, A.~Perelygin, J.~Wu, J.~Chuang, C.~D. Manning, A.~Y. Ng, and C.~Potts, ``Recursive deep models for semantic compositionality over a sentiment treebank,'' in \emph{Intern. Conf. on Empirical Methods in Natural Language Processing}, 2013, pp. 1631--1642.

\bibitem{mrpc}
B.~Dolan and C.~Brockett, ``Automatically constructing a corpus of sentential paraphrases,'' in \emph{Intern. Workshop on Paraphrasing (IWP2005)}, 2005.

\bibitem{rte}
L.~Bentivogli, P.~Clark, I.~Dagan, and D.~Giampiccolo, ``The fifth pascal recognizing textual entailment challenge.'' in \emph{Text Analysis Conference (TAC)}, 2009.

\bibitem{ibert}
S.~Kim, A.~Gholami, Z.~Yao, M.~W. Mahoney, and K.~Keutzer, ``{I-BERT}: Integer-only bert quantization,'' in \emph{Proc. of Machine Learning Research (PMLR)}, 18--24 Jul 2021, pp. 5506--5518.

\bibitem{gelu}
D.~Hendrycks and K.~Gimpel, ``Gaussian error linear units (gelus),'' \emph{arXiv preprint arXiv:1606.08415}, 2016.

\bibitem{online-softmax}
M.~Milakov and N.~Gimelshein, ``Online normalizer calculation for softmax,'' \emph{arXiv preprint arXiv:1805.02867}, 2018.

\bibitem{benini-islped23}
G.~Islamoglu, M.~Scherer, G.~Paulin, T.~Fischer, V.~J. Jung, A.~Garofalo, and L.~Benini, ``Ita: An energy-efficient attention and softmax accelerator for quantized transformers,'' in \emph{{IEEE} Intern. Symp. on Low Power Electronics and Design (ISLPED)}, 2023.

\bibitem{zhu2020efficient}
D.~Zhu, S.~Lu, M.~Wang, J.~Lin, and Z.~Wang, ``Efficient precision-adjustable architecture for softmax function in deep learning,'' \emph{Transactions on Circuits and Systems II: Express Briefs}, vol.~67, pp. 3382--3386, 2020.

\bibitem{python-pwl}
\BIBentryALTinterwordspacing
C.~F. Jekel and G.~Venter. (2019) {PWLF:} a python library for fitting 1d continuous piecewise linear functions. [Online]. Available: \url{https://github.com/cjekel/piecewise\_linear\_fit\_py}
\BIBentrySTDinterwordspacing

\bibitem{logarithm-kaist}
H.~Kim, B.-G. Nam, J.-H. Sohn, J.-H. Woo, and H.-J. Yoo, ``A 231-mhz, 2.18-mw 32-bit logarithmic arithmetic unit for fixed-point 3-d graphics system,'' \emph{{IEEE} journal of solid-state circuits}, vol.~41, no.~11, pp. 2373--2381, 2006.

\bibitem{glue}
A.~Wang, A.~Singh, J.~Michael, F.~Hill, O.~Levy, and S.~R. Bowman, ``{GLUE}: A multi-task benchmark and analysis platform for natural language understanding,'' in \emph{Intern. Conf. on Learning Representations (ICLR)}, 2019.

\bibitem{git}
\BIBentryALTinterwordspacing
C.~Peltekis, K.~Alexandridis, and G.~Dimitrakopoulos. (2024) {SIMD}-{S}oftmax. [Online]. Available: \url{https://github.com/ic-lab-duth/SIMD-Softmax.git}
\BIBentrySTDinterwordspacing

\end{thebibliography}

\end{document}